\documentclass{iaus}
\usepackage{graphicx}

\title[Stellar differential rotation and meridional flow] 
{Theory of differential rotation and meridional circulation}

\author[L.\,L.~Kitchatinov]   
{Leonid L. Kitchatinov}

\affiliation{
 Institute for Solar-Terrestrial Physics, \\
 Lermontov Str. 126a, PO Box 291, Irkutsk 664033, Russia \\
 email: {\tt kit@iszf.irk.ru} \\[\affilskip]
 Pulkovo Astronomical Observatory, \\
 St. Petersburg 196140, Russia
}

 \pubyear{2013}
 \volume{294}  
 \pagerange{XXX--XXX}
 \jname{Solar and Astrophysical Dynamos and Magnetic Activity}
 \editors{A.\,G.~Kosovichev, E.\,M.~de\,Gouveia\,Dal\,Pino \& Y.~Yan, eds.}
 
\begin{document}

\maketitle


\begin{abstract}
Meridional flow results from slight deviations from the thermal wind
balance. The deviations are relatively large in the boundary layers
near the top and bottom of the convection zone. Accordingly, the
meridional flow attains its largest velocities at the boundaries and
decreases inside the convection zone. The thickness of the boundary
layers, where meridional flow is concentrated, decreases with
rotation rate, so that an advection-dominated regime of dynamos is
not probable in rapidly rotating stars. Angular momentum transport
by convection and by the meridional flow produce differential
rotation. The convective fluxes of angular momentum point radially
inward in the case of slow rotation but change their direction to
equatorward and parallel to the rotation axis as the rotation rate
increases. The differential rotation of main-sequence dwarfs is
predicted to vary mildly with rotation rate but increase strongly
with stellar surface temperature. The significance of differential
rotation for dynamos has the opposite tendency to increase with
spectral type.
\end{abstract}

\firstsection 

 \section{Introduction}
Differential rotation and meridional flow are the two main
components of global stellar circulation. Both are important for
dynamos. Differential rotation winds toroidal magnetic fields, and
meridional flow near the base of the convection zone is probably
responsible for the observed equatorward migration of sunspot
activity. Knowledge of differential rotation as a function of
stellar parameters is a key for understanding stellar dynamos.

Helioseismology has shown that the regions inside the sun occupied
by thermal convection and differential rotation coincide. The same
is probably true of the meridional flow. Any theory of differential
rotation and meridional flow has to describe the global flows
against the background of convective turbulence. The large-scale and
turbulent flows are intimately linked. The theory, therefore, has to
rely on the tools and methods of the mean-field hydrodynamics of
turbulent fluids.

This paper reviews the mean-field theory of meridional flow and
differential rotation. The differential rotation can be understood
as a result of interaction between convection and rotation. The
meridional flow is produced by the non-conservative part of
centrifugal force and by the baroclinic torque provided that these
two drivers do not balance each other. Numerical models based on the
theory are also discussed. The models reproduce closely solar
rotation and predict the dependence of differential rotation on
stellar parameters. The results of 3D numerical simulations are
discussed only occasionally. Numerical experiments have been
reviewed recently by \cite[Miesch \& Toomre (2009)]{MT09} and
\cite[Brun \& Rempel (2009)]{BR09}.
 \section{Meridional circulation}\label{MF}
Meridional circulation is a vortical flow. Accordingly, the flow is
convenient to describe in terms of the azimuthal vorticity $\omega =
(\mbox{\boldmath $\nabla$}\times\mbox{\boldmath
$V$}^\mathrm{m})_\phi$, where $\mbox{\boldmath $V$}^\mathrm{m}$ is
the global (azimuthally-averaged) meridional velocity. The
mean-field vorticity equation,
 \begin{equation}
    \frac{\partial\omega}{\partial t} + r \sin\theta\ \mbox{\boldmath $\nabla$}\cdot
    \left(\mbox{\boldmath $V$}^\mathrm{m} \frac{\omega}{r\sin\theta}\right) +
    {\cal D}\left(\mbox{\boldmath $V$}^\mathrm{m}\right) =\ \sin\theta\ r{\partial\Omega^2\over\partial z}\
    -\ {g\over c_{\rm p} r}{\partial S\over\partial\theta},
    \label{1}
 \end{equation}
collects the meridional flow drivers on its right side while the
left side describes the flow reaction to this driving. In
particular, the term ${\cal D}\left(\mbox{\boldmath
$V$}^\mathrm{m}\right)$ accounts for the viscous resistance by the
eddy viscosity to the meridional flow driving,
 \begin{equation}
    {\cal D}(\mbox{\boldmath $V$}^\mathrm{m})\ =-\varepsilon_{\phi jk} \frac{\partial}{\partial r_j}
    \left(\frac{1}{\rho}\frac{\partial}{\partial r_l}
    \left(\rho\ {\cal N}_{klin}\frac{\partial V_i^\mathrm{m}}{\partial r_n}
    \right)\right) ,
    \label{2}
 \end{equation}
where ${\cal N}_{klin}$ is the eddy viscosity tensor and repetition
of subscripts signifies summation. In Eq.~(\ref{1}), the usual
spherical coordinates $(r,\theta ,\phi )$ are used, $\Omega$ is the
angular velocity, $S$ is the specific entropy, $g$ is gravity,
$c_\mathrm{p}$ is the specific heat at constant pressure and
$\partial /\partial z = \cos\theta \partial /\partial r -
r^{-1}\sin\theta \partial /\partial\theta$ is the spatial derivative
along the rotation axis.

\begin{figure}[htb]
\begin{center}
 \includegraphics[width = 9 truecm]{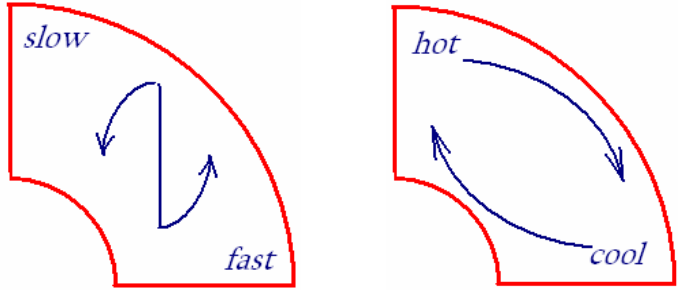}
 \caption{Illustration of centrifugal ({\sl left}) and
        baroclinic ({\sl right}) driving of the meridional
        flow (see text).}
   \label{f1}
\end{center}
\end{figure}

In hydrodynamics, there are only two main sources of meridional flow
illustrated by Fig.\,\ref{f1}. The first term on the right side of
Eq.~(\ref{1}) shows that differential rotation can produce the flow
(\cite[Kippenhahn 1963]{K63}). If angular velocity varies with the
cylindrical coordinate $z$ along the rotation axis, the centrifugal
force is not conservative and generates vorticity. This effect was
recently named \lq gyroscopic pumping' (\cite[Garaud \& Bodenheimer
2010]{GB10}). If angular velocity decreases with distance from the
equatorial plane, as it does in the sun, the centrifugal force
produces a torque driving anti-clockwise circulation (in the
north-west quadrant of the meridional cross-section).

The second term on the right side of Eq.~(\ref{1}) allows for the
baroclinic driving of the meridional flow. If polar regions are
warmer than the equator, as they probably are on the sun (\cite[Rast
et al. 2008]{ROM08}), then the warm fluid tends to rise up near
poles and spread over the surface while the cooler equatorial fluid
tends to sink down and spread over near the base of the convection
zone thus producing clockwise circulation (Fig.\,\ref{f1}).

The two sources of the meridional flow are competing on the sun.
This is probably not by chance. It seems to be a general rule that
these two sources are nearly balancing each other in convective
stars. If Eq.~(\ref{1}) is normalized to dimensionless units by
multiplying this equation by the square of the viscous diffusion
time, $R^4/\nu_{_\mathrm{T}}^2$, the first term on the right side
gets a coefficient of the Taylor number, $\mathrm{Ta} = 4\Omega^2
R^4/\nu^2_{_\mathrm{T}}$; $R$ is the stellar radius and
$\nu_{_\mathrm{T}}$ is the eddy viscosity. This number is large on
the sun, $\mathrm{Ta} \sim 10^7$. The second (baroclinic) term also
gets a big coefficient of the Grashof number, $\mathrm{Gr} =
(gR^3/\nu^2_{_\mathrm{T}})(\delta T/T) \sim 10^7 \delta
T_\mathrm{surf}$; $\delta T$ is the pole-equator temperature
difference and $\delta T_\mathrm{surf}$ is the surface value of this
differential temperature. The left side of the equation, on the
contrary, scales with the much smaller Reynolds number $\mathrm{Re}
= V_0 R/ \nu_{_\mathrm{T}} \sim 10$, where $V_0 \sim 10$~m/s is the
characteristic amplitude of the meridional flow (the second term on
the left side of (\ref{1}) scales with $\mathrm{Re}^2$). Therefore,
the Eq.~(\ref{1}) has to be satisfied by the two terms on the right
side nearly balancing each other.

Neglecting the left side in Eq.~(\ref{1}) gives the famous equation
of thermal wind balance
\begin{equation}
    0\ = \ \sin\theta\ r{\partial\Omega^2\over\partial z}\
    -\ {g\over c_{\rm p} r}{\partial S\over\partial\theta} .
    \label{33}
\end{equation}
The equation shows in particular that a positive differential
temperature of the order of 1\,K is necessary for isorotational
surfaces to deviate considerably from cylinders in solar rotation
models.

Equation (\ref{33}) shows what the thermal wind balance is but it
does not explain how the balance is maintained. In order to
understand this, we have to return to the meridional flow equation
(\ref{1}) and imagine that the balance is somehow violated. This
gives a strong source of meridional flow. The excited flow reacts
back on the distributions of angular velocity and entropy by
transporting angular momentum and heat to reestablish the balance.
Meridional flow maintains the thermal wind balance and deviations
from the balance drive the flow.

\begin{figure}[htb]
\begin{center}
 \includegraphics[width = 5.1 truecm]{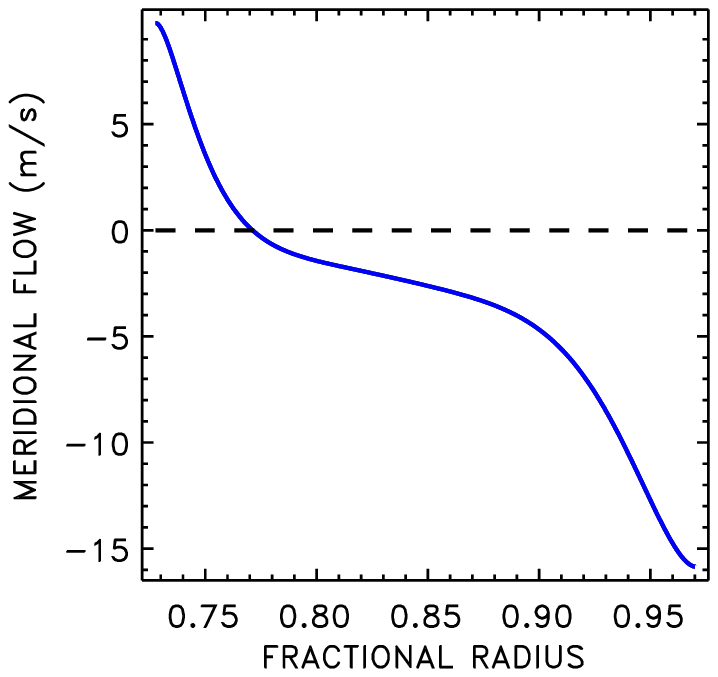}
 \hspace{.5truecm}
 \includegraphics[width = 5.1 truecm]{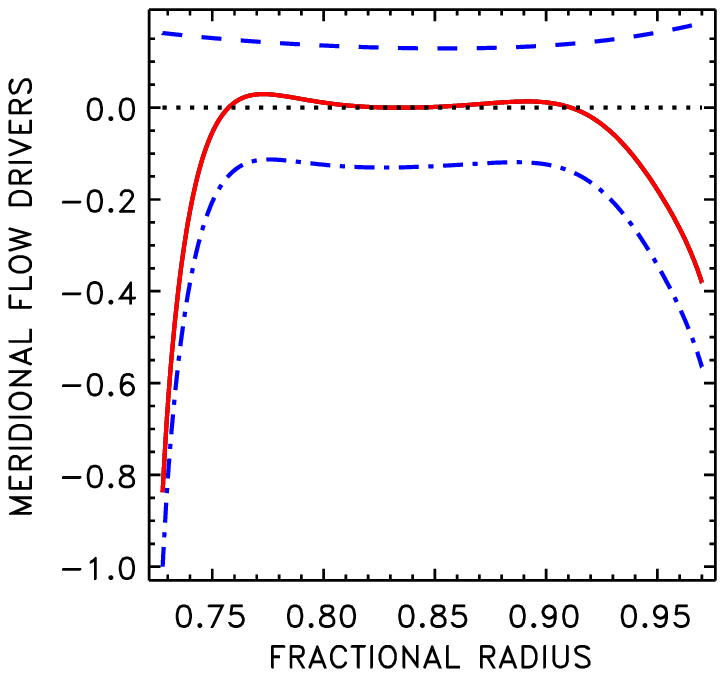}
 \caption{{\sl Left:} depth profile of the meridional velocity for
     45$^\circ$ latitude after the mean-field model of
     \cite[Kitchatinov \& Olemskoy (2011)]{KO11}.
     Negative velocity means poleward flow.
     {\sl Right:} dependencies of the baroclinic (dashed) and
     centrifugal (dashed-dotted) terms of Eq.\,(\ref{1}) and their
     sum (full line) on depth for the same latitude. The meridional
     flow attains its largest velocities in the boundary layers
     where the thermal wind balance is violated.
 }
   \label{f2}
\end{center}
\end{figure}

This qualitative picture can explain why the meridional flow
decreases with depth beneath the solar surface (\cite[Zhao \&
Kosovichev 2004]{ZK04}). It does so because the stress-free
conditions on the boundaries of the convection zone are not
compatible with thermal wind balance. Accordingly, the boundary
layers form where the balance is violated (\cite[Durney 1989]{D89}).
As deviations from the balance produce the meridional flow, the
meridional velocity attains its maximum values at the top and bottom
boundaries and decreases inside the convection zone. Fig.~\ref{f2}
illustrates the thermal wind balance and the meridional flow
structure typical of mean-field models of global solar circulation.
Boundary layers were found also in the 3D simulations of \cite[Brown
et al. (2008)]{Bea08}. The bottom flow of Fig.\,\ref{f2} is not
small compared to the surface flow. The flow, however, decreases
rapidly with depth beneath the base of the convection zone
(\cite[Gilman \& Miesch 2004]{GM04}).

Fig.\,\ref{f2} shows the boundary layers where the thermal wind
balance is violated. The bulk of the convection zone is,
nevertheless, very close to the balance. Even there, however, the
balance condition is satisfied only on average (Fig.\,\ref{f2}
results from the mean-field model). Differential rotation is
produced by turbulent convection, so that fluctuating deviations
from the thermal wind balance are unavoidable (\cite[Brun et al.
2010]{BAC10}). As the deviations produce the meridional flow, the
flow is expected to fluctuate considerably with time. This
qualitative picture was reproduced quantitatively by \cite[Rempel
(2005)]{R05} whose global circulation model allowed for random
fluctuations in the $\Lambda$-effect (cf. the next Section for the
$\Lambda$-effect definition). The fluctuating $\Lambda$-effect
produces small fluctuations in differential rotation, which in turn
produce much larger irregular variations in the meridional flow. The
relative amplitude of fluctuations of the meridional flow in
Rempel's model was about two orders of magnitude larger compared to
fluctuations in angular velocity. Helioseismologically detected
meridional flow indeed shows considerable changes from year to year
(\cite[Zhao \& Kosovichev 2004]{ZK04}; \cite[Gonz\'ales Hern\'andez
et al. 2006]{Gea06}) contrasting the slight variations in the
rotation rate.
 \section{Differential rotation}
Helioseismology shows that differential rotation and thermal
convection occupy the same region inside the sun.  The entire
convection zone is rotating differentially while the rotation
inhomogeneity decreases rapidly with depth beneath the convection
zone (\cite[Wilson et al. 1997]{WBL97}; \cite[Schou et al.
1998]{Sea98}). This supports the theoretical concept pioneered by
\cite[Lebedinskii (1941)]{L41} that the differential rotation
results from interaction between convection and rotation. Convection
in a rotating fluid is disturbed by the Coriolis force. The back
reaction disturbs rotation to make it non-uniform.

The mean-field equation for angular velocity,
\begin{equation}
    \rho r^2\sin^2\theta \frac{\partial\Omega}{\partial t} =
    -\mbox{\boldmath $\nabla$}\cdot\left( \rho r\sin\theta \langle
    u_\phi \mbox{\boldmath$u$}\rangle + \rho  r^2\sin^2\theta\
    \Omega\mbox{\boldmath$V$}^\mathrm{m}\right) ,
    \label{3}
\end{equation}
shows that the angular velocity distribution can be modified by
angular momentum transport, by convection ({\boldmath$u$}) and by
meridional flow ({\boldmath$V$}$^\mathrm{m}$).
 \subsection{The $\Lambda$-effect}
The ability of convective motions to transport angular momentum even
in the case of uniform rotation is called the \lq $\Lambda$-effect'
(\cite[R\"udiger 1989]{R89}). The uniformity of rotation is
mentioned because the eddy viscosity can transport angular momentum
also if rotation is not homogeneous. The viscosity, however, tends
to diminish the differential rotation, not to produce it, in
contrast with the non-diffusive $\Lambda$-effect.

Eq.~(\ref{3}) shows that the azimuthal and meridional convective
velocities have to be correlated for the $\Lambda$-effect to emerge.
The finite cross-correlation requires the convective turbulence to
possess a preferred direction of anisotropy or inhomogeneity (cf.
\cite[Kitchatinov (2011)]{K11} for pictorial discussion of the
origin of the $\Lambda$-effect). The radial preferred direction is
imposed by gravity. The $\Lambda$-effect is allowed for by the
non-diffusive part $Q_{ij}^\Lambda$ of the velocity correlation
tensor $Q_{ij} = \langle u_i u_j\rangle$. The radial
($Q^\Lambda_{r\phi}$) and meridional ($Q^\Lambda_{\theta\phi}$)
non-diffusive fluxes of angular momentum for radially-stratified
fluid read
\begin{eqnarray}
    Q^\Lambda_{r\phi} &=&-\nu_{_\mathrm{T}}
    \left(\frac{\ell}{H_\rho}\right)^2\Omega\sin\theta
    \left( V(\Omega^*) + H(\Omega^*)\cos^2\theta\right) ,
    \nonumber \\
    Q^\Lambda_{\theta\phi} &=& \nu_{_\mathrm{T}}
    \left(\frac{\ell}{H_\rho}\right)^2\Omega\sin^2\theta
    \cos\theta\ H(\Omega^*).
    \label{4}
\end{eqnarray}
where $\ell$ is the correlation length,
\begin{equation}
    \nu_{_\mathrm{T}} = - \frac{\tau\ell^2 g}{15 c_\mathrm{p}}
    \frac{\partial S}{\partial r}
    \label{5}
\end{equation}
is the background eddy viscosity, and $\tau$ is the convective
turnover time. The dimensionless functions $V(\Omega^*)$ and
$H(\Omega^*)$ in Eq.~(\ref{4}) depend on the Coriolis number
\begin{equation}
    \Omega^* = 2\tau\Omega ,
    \label{6}
\end{equation}
which is the key parameter of the differential rotation theory. The
parameter measures intensity of interaction between convection and
rotation. Its value defines whether the convective eddies are living
long enough for rotation to influence them considerably. The
Coriolis number (\ref{6}) depends on depth in the convection zone.
It is smaller than one near the solar surface but increases with
depth to exceed ten near the base of the convection zone. Therefore,
the $\Lambda$-effect theory should be nonlinear in $\Omega^*$.

\begin{figure}[htb]
\begin{center}
 \includegraphics[height = 4.7 truecm]{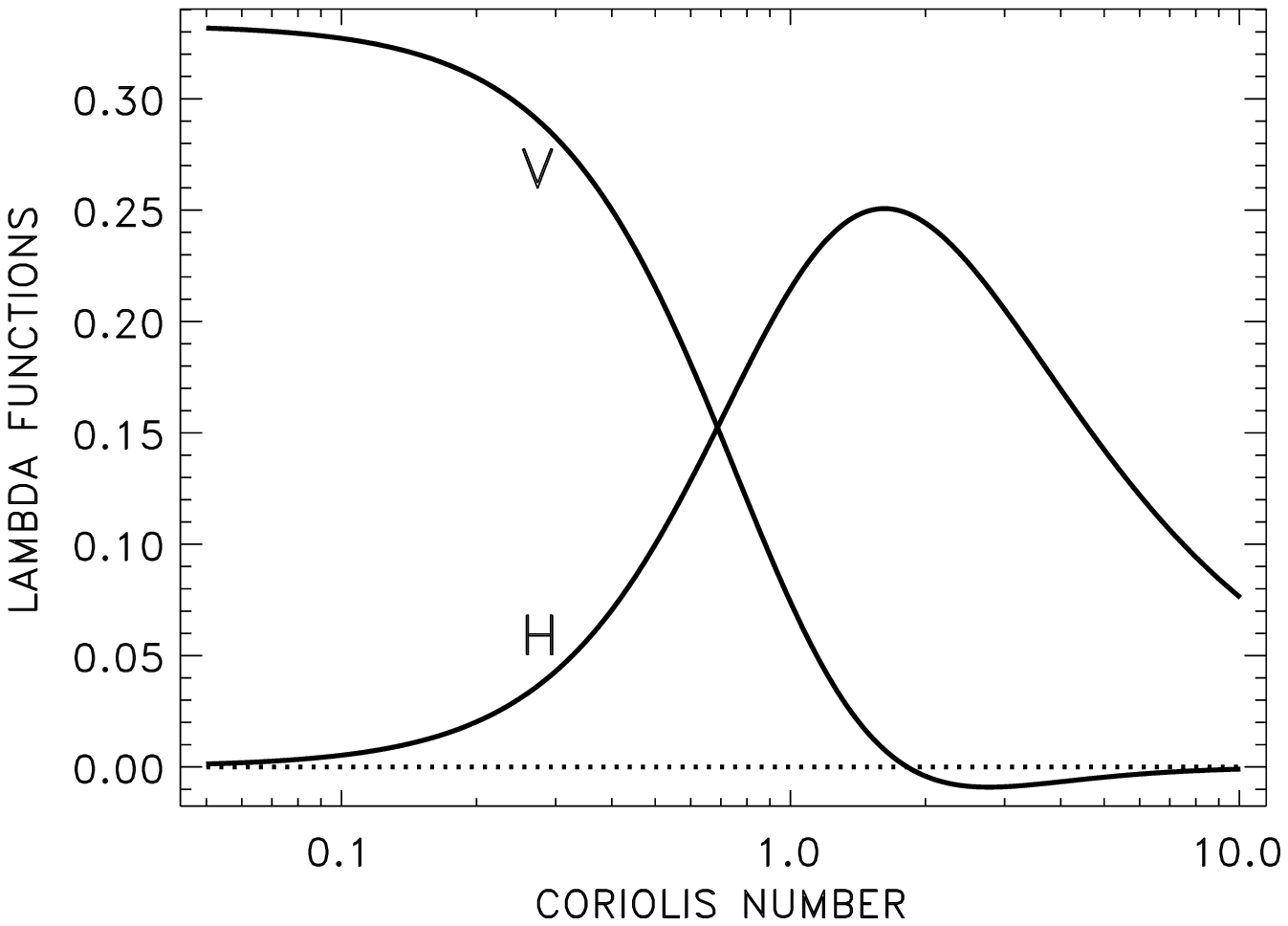}
 \hspace{1.0truecm}
 \includegraphics[height = 4.7 truecm]{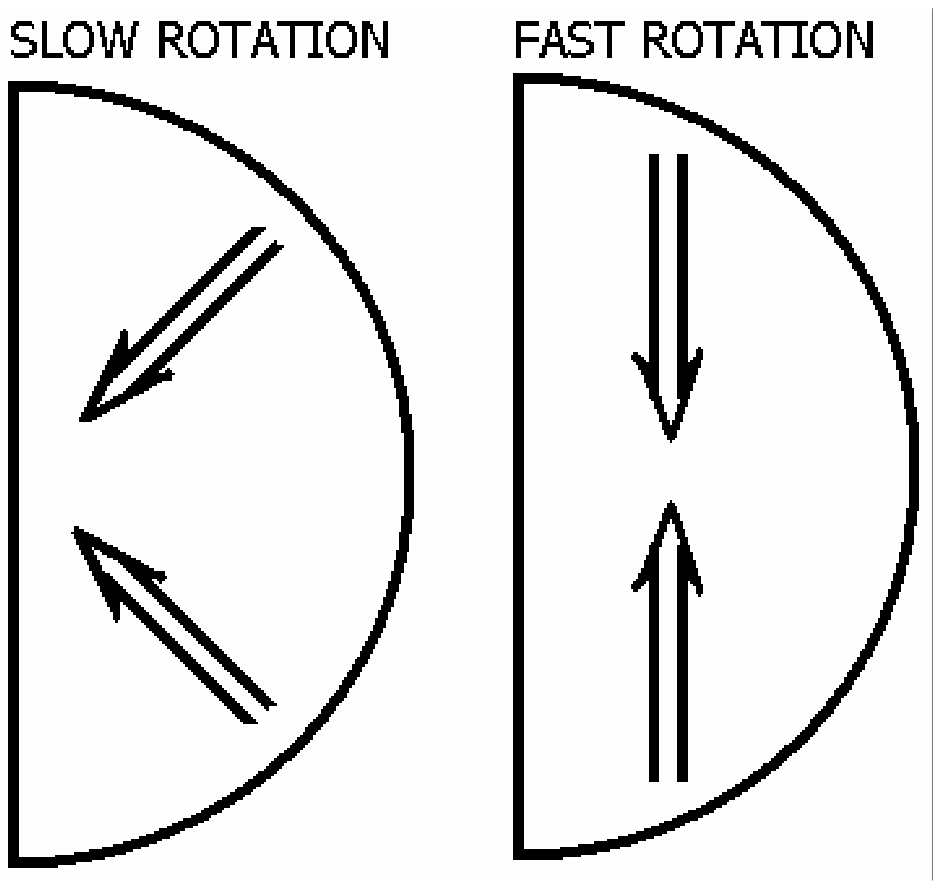}
 \caption{The functions $V(\Omega^*)$ and $H(\Omega^*)$ of the
    $\Lambda$-effect of Eq.~(\ref{4}) derived by
    \cite[Kitchatinov \& R\"udiger (1993, 2005)]{KR93;KR05}. The
    right panel shows the directions of angular momentum transport
    for the cases of slow ($\Omega^* \ll 1$) and fast ($\Omega^* \gg
    1$) rotation.
 }
   \label{f3}
\end{center}
\end{figure}

The function $V(\Omega^*)$ in Eq.\,(\ref{4}) is the normalized
vertical flux of angular momentum. Positive $V$ means radial inward
transport. The function $H(\Omega^*)$ is the normalized flux
parallel to the rotation axis. Its positive value means that the
angular momentum is transported towards the equatorial plane.  The
two (not mutually orthogonal) components of the angular momentum
fluxes are shown in Fig.~\ref{f3} as functions of the Coriolis
number. In the case of slow rotation, $\Omega^* \ll 1$, the
$H$-function is relatively small and angular momentum is transported
downward in radius. In the opposite case of rapid rotation,
$\Omega^* \gg 1$, the $V$-function is small, and the angular
momentum flux is parallel to the rotation axis pointing to the
equatorial plane. As the Coriolis number increases from small to
large values, the direction of the non-diffusive flux of angular
momentum varies smoothly from radial inward to equatorward and
parallel to the rotation axis.
 \subsection{Eddy viscosities}
After the significance of the correlation of azimuthal and
meridional convective velocities for differential rotation was
recognized, there were attempts to observe this correlation using
sunspots as tracers (\cite[Ward 1965]{W65}). The first measurements
gave finite correlation, $Q_{\theta\phi}\cos\theta \sim
10^3$\,m$^2$/s$^2$, with positive $Q_{\theta\phi}$ in the northern
and negative in the southern hemisphere, indicating the equatorward
transport of angular momentum. Subsequent measurements of
\cite[Ribes (1986)]{R86} and \cite[Nesme-Ribes et al.
(1993)]{NRea93}, however, did not confirm this result.

The equatorial acceleration does not demand a certain sign of
$Q_{\theta\phi}$ (a positive value of $Q_{\theta\phi}\cos\theta$
would mean that angular momentum is permanently pumped towards the
equator implying an increase of differential rotation with time).
The total correlation,
\begin{equation}
    Q_{ij} = Q^\Lambda_{ij} + Q^\nu_{ij} ,
    \label{7}
\end{equation}
includes the contribution $Q^\nu$ of the eddy viscosities in line
with the $\Lambda$-effect. The viscous fluxes of angular momentum,
\begin{eqnarray}
    Q_{r\phi}^\nu &=& - \nu_1 r \sin\theta
    \frac{\partial\Omega}{\partial r}-\nu_2 \sin\theta
    \cos\theta \left( r \cos\theta \frac{\partial \Omega}{\partial r} -
    \sin\theta \frac{\partial\Omega}{\partial\theta}\right) ,
    \nonumber\\
    Q_{\theta\phi}^\nu &=& -
    \nu_1  \sin\theta \frac{\partial \Omega}
    {\partial \theta} + \nu_2 \sin^2\theta
    \left( r \cos\theta \frac{\partial \Omega}{\partial r} -
    \sin\theta \frac{\partial\Omega}{\partial\theta}\right)
    \label{8}
\end{eqnarray}
(\cite[Kitchatinov et al. 1994]{Kea94}), include two viscosity
coefficients, $\nu_1 = \nu_{_\mathrm{T}}f_1(\Omega^*)$ and $\nu_2 =
\nu_{_\mathrm{T}}f_2(\Omega^*)$. Rotating convection is anisotropic
and the eddy viscosity $\nu_1$ acting in the direction normal to the
rotation axis differs from the viscosity $\nu_1 + \nu_2$ for the
direction parallel to the axis.

A  positive non-diffusive part, $Q^\Lambda_{\phi\theta}\cos\theta >
0$, is required for maintaining equatorial acceleration. It is
hardly possible, however, to separate the contributions of the
$\Lambda$-effect from the contribution of the eddy viscosities in
the total correlation (\ref{7}) supplied by observations. The same
separation problem is met by 3D numerical simulations of the
$\Lambda$-effect. It was, nevertheless, possible to extract the
$Q^\Lambda$ from the numerical simulations by \cite[K\"apyl\"a \&
Brandenburg (2008)]{KB08} and \cite[K\"apyl\"a et al.
(2011)]{Kea11}. The analytical results in Fig.~\ref{f3} generally
agree with the simulations of the $\Lambda$-effect.
 \subsection{Surface shear layer}
The natural boundary condition for the differential rotation problem
is the requirement that the surface density of external forces be
zero, $Q_{r\phi} = Q_{r\theta} = 0$. The stress-free condition means
that the global flows are produced by internal processes in the
convection zone, not imposed externally. The condition on the upper
boundary is simplified by the small value of the Coriolis number so
that the angular momentum fluxes (\ref{4}) and (\ref{8}) can be
taken in the limit of $\Omega^* \rightarrow 0$. In this limit, the
viscosity $\nu_2$ in Eq.~(\ref{8}) vanishes and $\nu_1 =
\nu_{_\mathrm{T}}$. The top boundary condition, $Q_{r\phi} = 0$,
then reads
\begin{equation}
    r \frac{\partial\Omega}{\partial r} = -
    \left(\frac{\ell}{H_\rho}\right)^2 V(0) \Omega \ \ \ \mathrm{at}
    \ \ r = R .
    \label{9}
\end{equation}
With positive $V(0)$ of Fig.~\ref{f3}, Eq.~(\ref{9}) requires the
angular velocity to increase with depth near the surface in at least
qualitative agreement with the surface shear layer detected by
helioseismology (\cite[Schou et al. 1998]{Sea98}).

We have seen in Sect.~\ref{MF} that the meridional flow also has a
near-surface boundary layer. The near-surface rotational shear and
meridional flow are mutually related. This relation was recently
analyzed by \cite[Miesch \& Hindman (2011)]{MH11}.
 \section{Differential temperature}
It is not possible to reproduce the helioseismological rotation law
by differential rotation models if the differential temperature is
not allowed for (\cite[Brandenburg et al. 1990]{Bea90};
\cite[R\"udiger et al. 2005]{Rea05}; \cite[Miesch et al.
2006]{MBT06}).

\begin{figure}[htb]
\begin{center}
\includegraphics[width=7cm]{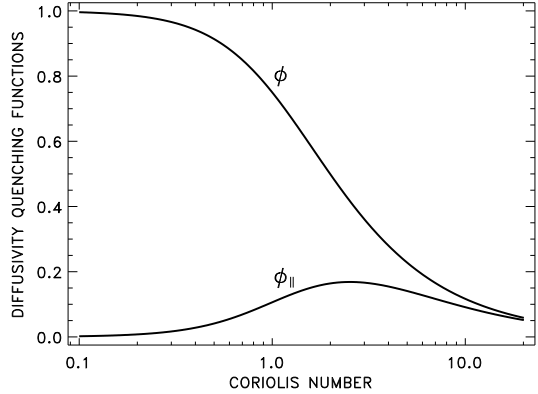}
         \caption{Quenching functions of the eddy thermal diffusivity of
         Eq.~(\ref{11}) after the quasilinear theory of turbulent
         transport in rotating fluids (Kitchatinov et al.\ 1994).
            }
         \label{f4}
\end{center}
\end{figure}

Similar to the differential rotation, the differential temperature
is currently understood as a result of rotational influence on
convection. The tensor $\chi_{ij}$ of the eddy thermal diffusion
that controls the convective heat flux,
\begin{equation}
    F^\mathrm{conv}_i = -\rho T \chi_{ij}
    \frac{\partial S}{\partial r_j} ,
    \label{10}
\end{equation}
includes the rotationally-induced anisotropy and quenching
\begin{equation}
    \chi_{ij} = \chi_{_\mathrm{T}} \left( \phi(\Omega^*) \delta_{ij}
    + \phi_\|(\Omega^*) \frac{\Omega_i\Omega_j}{\Omega^2}\right) , \ \ \ \
    \chi_{_\mathrm{T}} = -\frac{\tau\ell^2 g}{12 c_\mathrm{p}}
    \frac{\partial S}{\partial r}.
    \label{11}
\end{equation}
The diffusivity quenching functions of the Coriolis number (\ref{6})
are shown in Fig.~\ref{f4}.

Directions of the entropy gradient and heat flux (\ref{10}) do not
coincide due to the anisotropy of heat transport. Even if entropy
varies mainly with radius, the meridional flux,
\begin{equation}
    F^\mathrm{conv}_\theta = \rho T \chi_{_\mathrm{T}}
    \phi_\|(\Omega^*) \cos\theta \sin\theta \frac{\partial
    S}{\partial r} ,
    \label{12}
\end{equation}
does not vanish. For the negative radial gradient of entropy in
convection zones, the meridional flux (\ref{12}) transports heat to
the poles. There have been multiple attempts to observe differential
temperature on the sun. Recent observations of \cite[Rust et al.
(2008)]{Rea08} suggest that polar regions are warmer than the
equator by about 2.5\,K. After the Eq.~(\ref{33}) of thermal wind
balance, this small differential temperature suffices for the
isorotational surfaces to deviate considerably from cylinders.
 \section{Differential rotation models}
What is usually meant by the \lq differential rotation models' are
numerical solvers which define not only the angular velocity but
also meridional flow and entropy distributions in stellar convection
zones.
 \subsection{The Sun}
\begin{figure}[htb]
\begin{center}
\includegraphics[height = 4.5 cm]{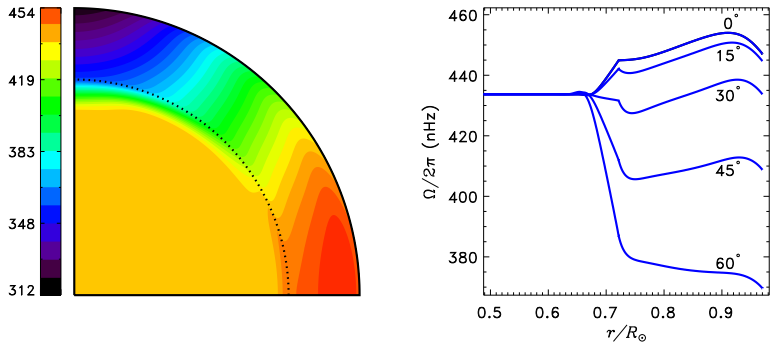}
         \caption{Angular velocity isolines ({\sl left}) and
         depth-dependencies of rotation frequencies for several
         latitudes ({\sl right}) after the mean-filed model of
         \cite[Kitchatinov \& Olemskoy (2011)]{KO11}.
            }
         \label{f5}
\end{center}
\end{figure}

The solar rotation model in Fig.~\ref{f5} is close to the
helioseismological rotation law. This Figure was computed by using
two different models - one for the convection zone, and another
model for the tachocline and radiation zone rotation. Physical
conditions in convection and radiation zones differ so much that it
is not possible to cover both in one model. The tachocline modeling
did not influence the computation of differential rotation in the
convection zone in any way but just used the results of this
computation as the top boundary condition.

The model of the convection zone rotation in Fig.~\ref{f5} was tuned
with one free parameter. Mean-field models can avoid arbitrary
prescriptions of any parameters. All the modeling needs, including
the $\Lambda$-effect, the eddy viscosities and thermal
conductivities, has been derived using the same quasi-linear
approximation of the mean-field theory. These derivations, formally,
leave no freedom for model design. However, the quasi-linear
approximation is not expected to be very precise. This raises the
question of how sensitive the model is to variations of basic
parameters. The sensitivity is normally quite low, except for one
parameter. E.g., if the $\Lambda$-effect is reduced by 50\% by
multiplying the non-diffusive fluxes of Eq.~(\ref{4}) by a factor of
0.5, the surface differential rotation in the solar model decreases
by only 10\%: from 30\% to 27\%. It is remarkable that increasing
the $\Lambda$-effect by 50\% (factor 1.5) also {\sl reduces} the
differential rotation by same 10\%. Sensitivity to an increase of
eddy viscosities or heat conductivities is also rather low
(decreasing the eddy diffusion can make the model unstable to
thermal convection and, therefore, inconsistent). The model,
however, is rather sensitive to the anisotropy of thermal
conductivity. If the parameter $C_\chi$ is introduced in
Eq.~(\ref{11}),
\begin{equation}
    \chi_{ij} = \chi_{_\mathrm{T}} \left( \phi(\Omega^*) \delta_{ij}
    + C_\chi\phi_\|(\Omega^*) \frac{\Omega_i\Omega_j}{\Omega^2}\right) ,
    \label{13}
\end{equation}
the resulting differential rotation reacts quite sensitively to
variations of this parameter and generally increases with $C_\chi$.
Fig.~\ref{f5} was obtained with $C_\chi = 1.5$. This implies that
the quasi-linear theory underestimates the anisotropy of thermal
eddy conductivity. All the stellar models discussed below were
obtained with the same value of $C_\chi = 1.5$.

Meridional flow for the same model as Fig.~\ref{f5} is shown in
Fig.~\ref{f2}.
 \subsection{Solar twins}
\begin{figure}[htb]
\begin{center}
 \includegraphics[height = 3.8 truecm]{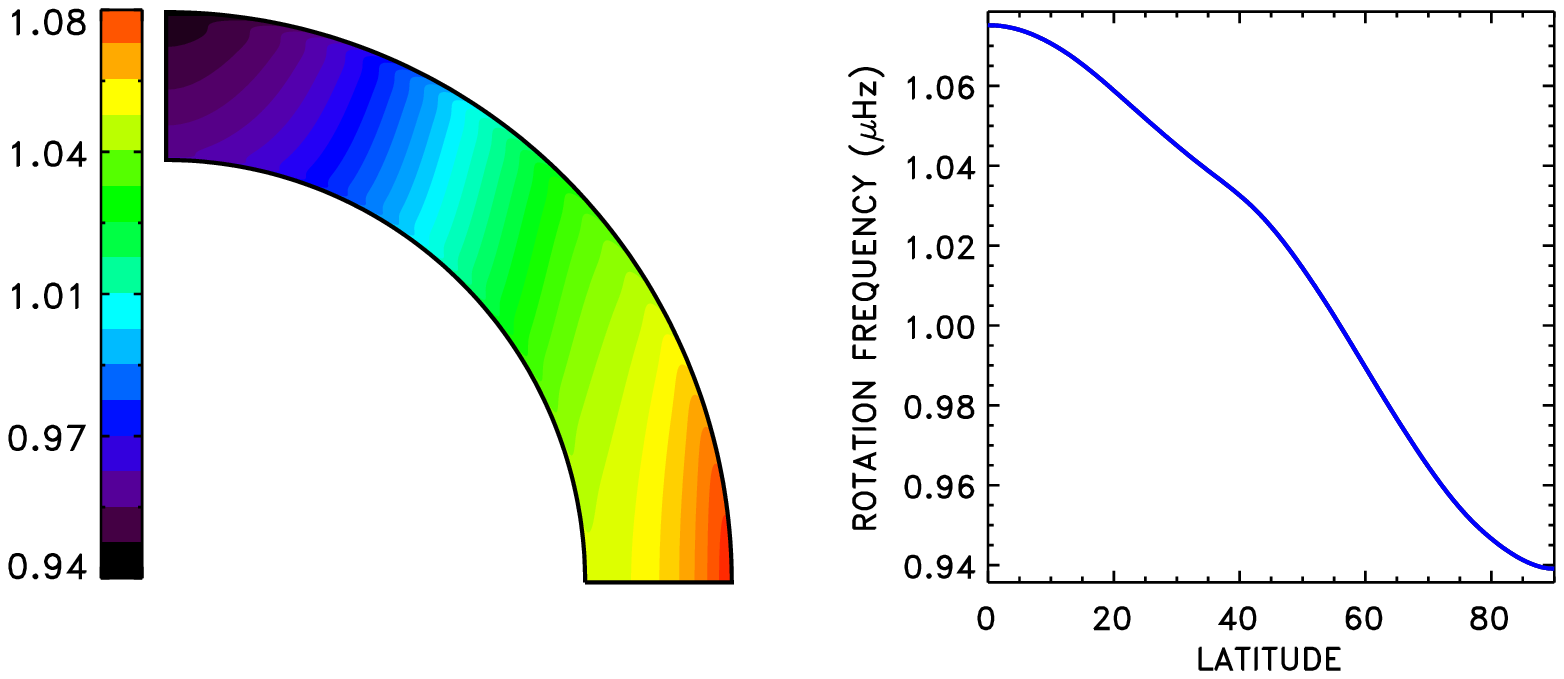}
 \hspace{0.2truecm}
 \includegraphics[height = 3.8 truecm]{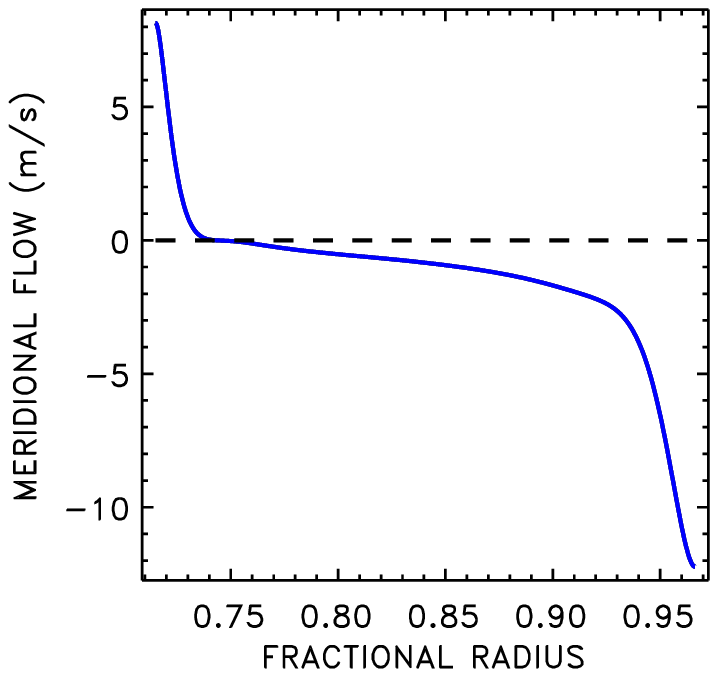}
 \caption{Differential rotation and meridional flow computed for
     solar-type star $\epsilon$\,Eridani. From left to right:
     angular velocity isolines, surface profile of rotation rate,
     and depth profile of the meridional flow at
     45$^\circ$-latitude.
 }
   \label{f6}
\end{center}
\end{figure}

\begin{figure}[htb]
\begin{center}
 \includegraphics[height = 3.8 truecm]{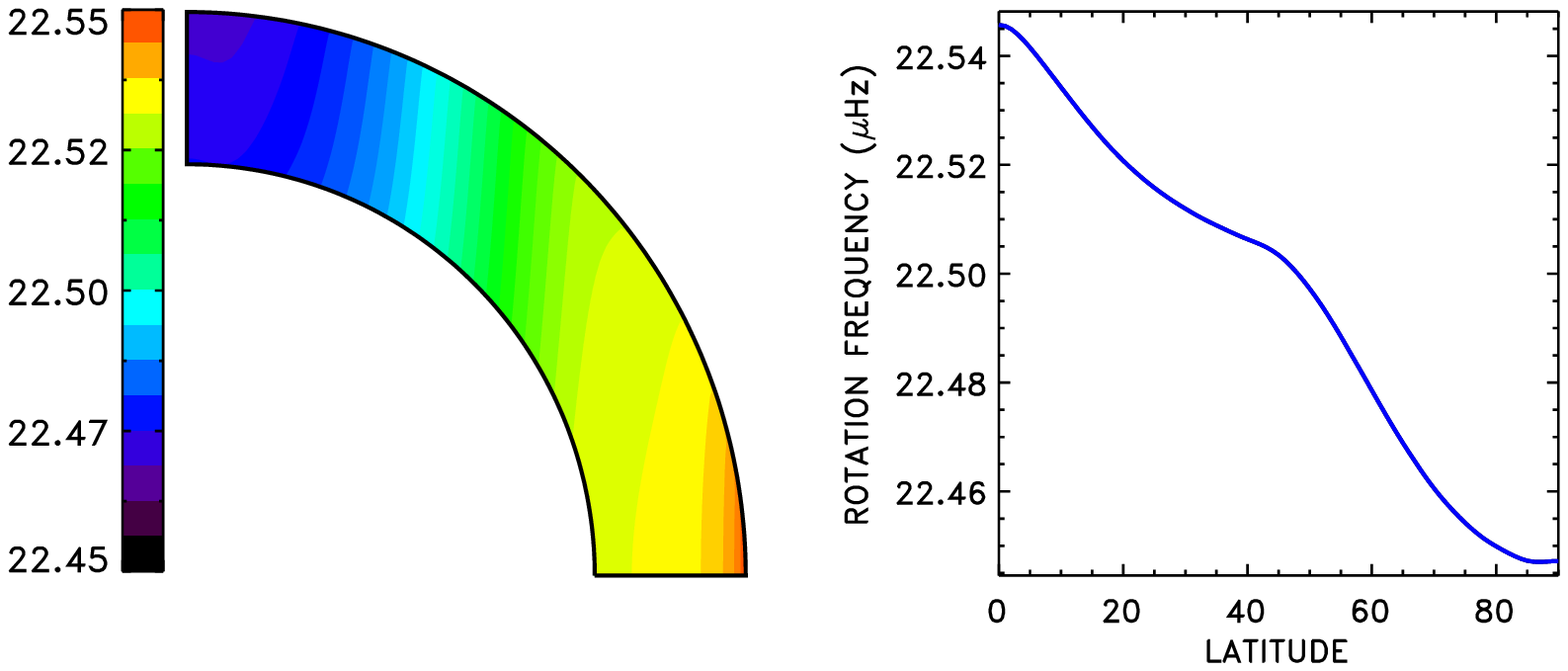}
 \hspace{0.2truecm}
 \includegraphics[height = 3.8 truecm]{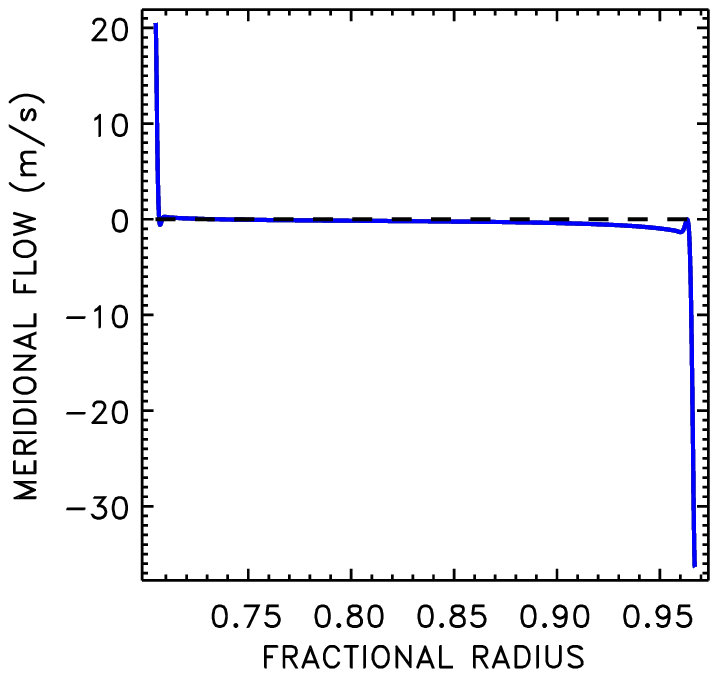}
 \caption{Differential rotation and meridional flow computed for
     the young rapidly rotating dwarf AB~Doradus. From left to right:
     angular velocity isolines, surface profile of rotation rate,
     and depth profile of the meridional flow at
     45$^\circ$-latitude.
 }
   \label{f7}
\end{center}
\end{figure}

Differential rotation has already been measured for many solar-type
stars (\cite[Korhonen 2012]{KH12}) motivating stellar applications
of differential rotation models. Figures~\ref{f6} and \ref{f7} show
the differential rotation and meridional flow computed for two stars
similar to the sun in mass but rotating faster (\cite[Kitchatinov \&
Olemskoy 2011]{KO11}).

\cite[Croll et al. (2006)]{Cea06} measured the differential rotation
of a solar twin $\epsilon$\,Eridani using very precise photometry of
the {\sl MOST} mission. The star is younger than the sun and rotates
with a shorter period of $P_\mathrm{rot} \simeq 11$~days. The
differential rotation and meridional flow computed for this star is
shown in Fig.~\ref{f6}. The computations give the relative
magnitude,
 \begin{equation}
    \alpha_\mathrm{DR} = 1 -
    \frac{\Omega_\mathrm{pole}}{\Omega_\mathrm{eq}} ,
    \label{14}
 \end{equation}
of the surface differential rotation of $\epsilon$\,Eri of
$\alpha_\mathrm{DR} = 0.13$ close to the value of
$\alpha_\mathrm{DR} = 0.11$ measured by \cite[Croll et al.
(2006)]{Cea06}. The absolute value of the surface differential
rotation is close to that of the sun.

Fig.~\ref{f7} shows differential rotation and meridional flow
computed for the very rapidly rotating ($P_\mathrm{rot} = 0.51$~day)
young dwarf AB~Doradus. The computed value of $\alpha_\mathrm{DR} =
4.4\times 10^{-3}$ is again close to the differential rotation
measured by \cite[Donati \& Collier Cameron (1997)]{DC97} by the
method of Doppler imaging. Also in this case the differential
rotation in dimensional units is close to the solar value.

The surface profiles of Figs.~\ref{f6} and \ref{f7} are not as
smooth as it is on the sun. There are peculiarities positioned
around the latitude, where the isoline of angular velocity
tangential to the inner boundary at the equator arrives on the
surface.

The meridional flows of these Figures differ from the flow computed
for the sun (Fig.~\ref{f2}) by their higher concentration at the
boundaries of the convection zone. The boundary layers discussed in
Sect.~2 become thinner as rotation rate increases. The flow of
Fig.~\ref{f7} in the very rapidly rotating star consists of two
near-boundary jets linked by very slow circulation in the bulk of
the convection zone. The meridional flow is believed important for
the solar dynamo (\cite[Choudhuri 2008]{C08}). It is, however, hard
to imagine that the surface-jets flow of Fig.~\ref{f7} can be
significant for dynamos. As a star ages and its rotation rate
decreases, the boundary layers broaden (Fig.~\ref{f6}) and the
meridional flow eventually transforms to a distributed flow as in
Fig.~\ref{f2}. It may be expected that the dynamo of a star changes
to the advection-dominated type from some other dynamo regime, where
meridional flow is not significant, as rotation rate decreases. This
may be the reason for the presence of two separate branches for slow
and fast rotators in the observed dependence of the stellar activity
cycle periods on rotation rate (\cite[Saar \& Brandenburg
1999]{SB99}).
 \subsection{Mass and rotation rate dependence}
\begin{figure}[htb]
\begin{center}
 \includegraphics[height = 5.1 truecm]{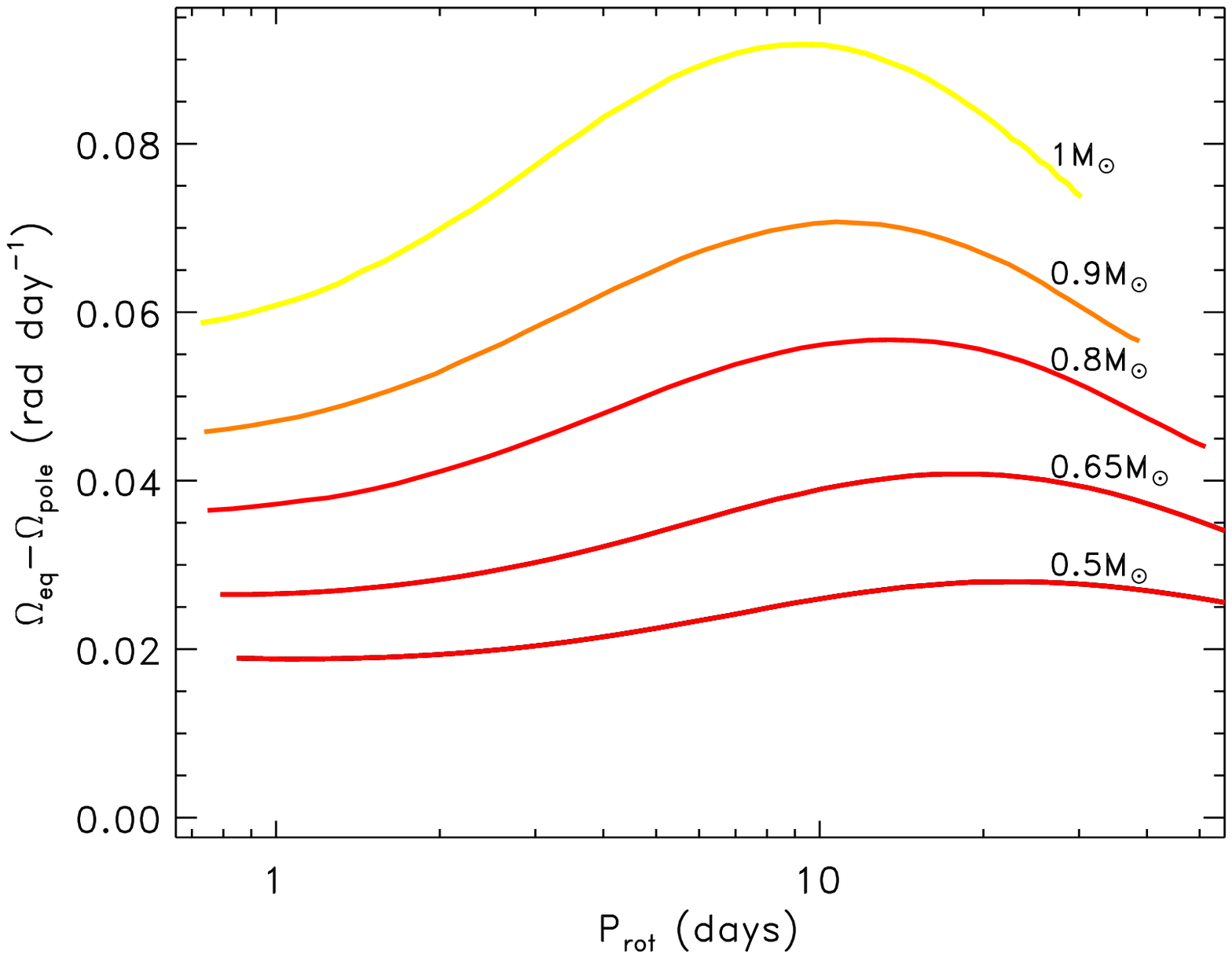}
 \hspace{0.2truecm}
 \includegraphics[height = 5.1 truecm]{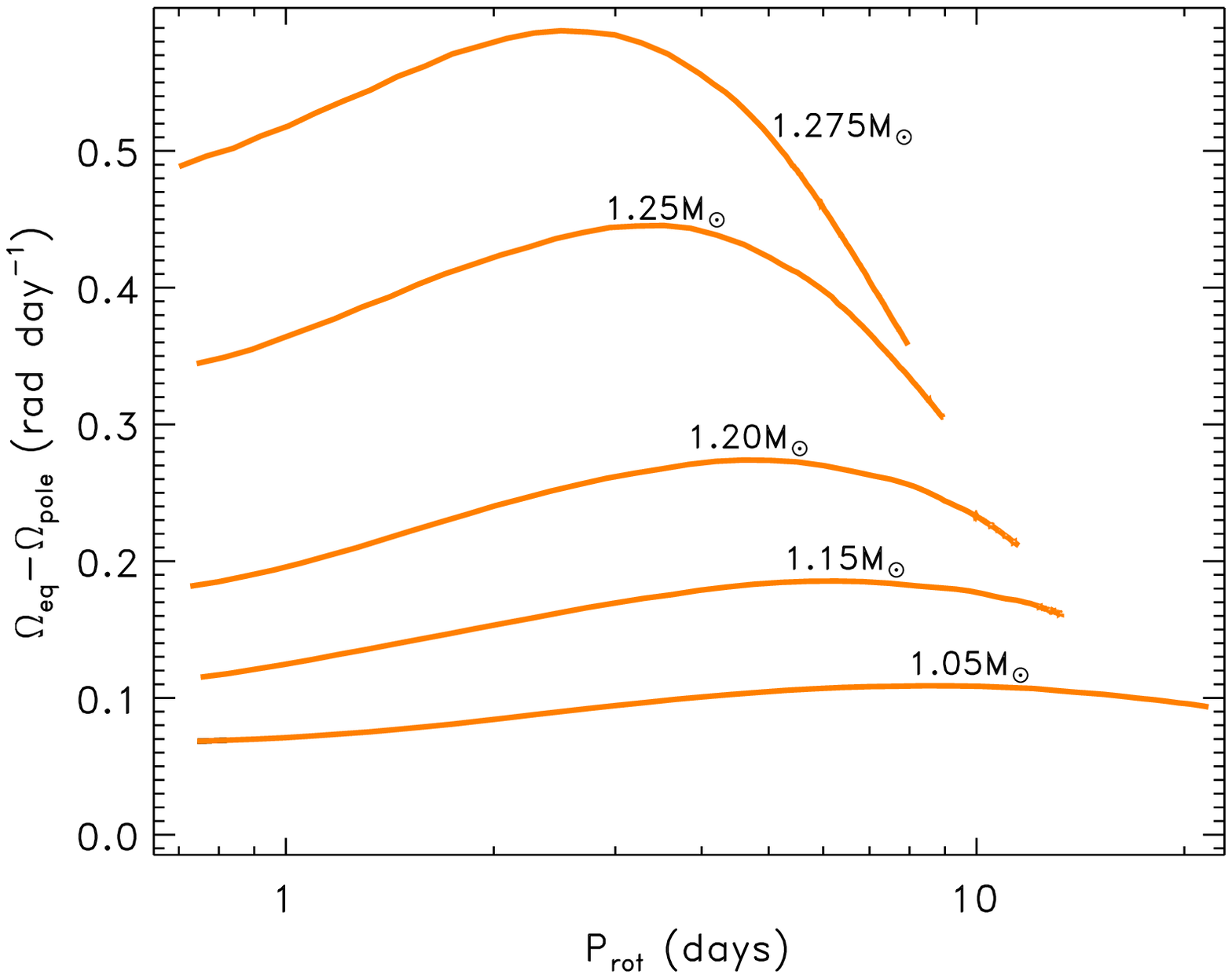}
 \caption{Surface differential rotation as a function of the rotation
     period for main-sequence dwarfs of subsolar ({\sl left}) and
     supersolar ({\sl right}) masses (from \cite[Kitchatinov \& Olemskoy 2012]{KO12}).
 }
   \label{f8}
\end{center}
\end{figure}

Applications of the mean-field model to individual stars do not
always agree with observations so closely as in the two examples
given above. However, no one case of clear disagreement has been met
so far. This motivates using the model to explore the dependence of
differential rotation on stellar parameters. Fig.~\ref{f8} shows the
surface differential rotation computed for main-sequence stars of
different mass and age. The rotation periods of this Figure were
specified using Gyrochronology (\cite[Barnes 2007, 2009]{B07}).

Fig.~\ref{f8} suggests that the differential rotation of a star of
given mass varies by about 30\% only as the star ages. A much
stronger tendency is the increase in differential rotation with
stellar mass. The same tendency for rapidly rotating stars was
observed by \cite[Barnes et al. (2005)]{Bea05}.

\begin{figure}[htb]
\begin{center}
 \includegraphics[height = 5.3 truecm]{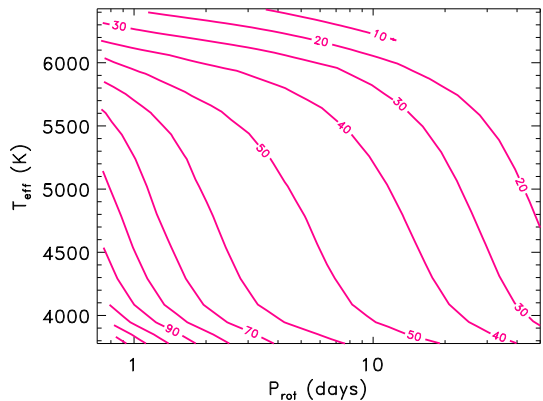}
 \caption{Isolines of the $C_\Omega$ dynamo number (\ref{15}) from the
    same computations as Fig.~\ref{f8}. The dynamo number increases
    with rotation rate but {\sl decreases} with temperature.
 }
   \label{f9}
\end{center}
\end{figure}

Differential rotation takes part in stellar dynamos by winding the
toroidal field lines. The computations in Fig.~\ref{f8} predict that
the largest differential rotation belongs to the hottest F-stars in
this Figure. The question arises whether the strong differential
rotation implies high dynamo activity. The computations suggest a
negative answer. The efficiency of differential rotation in
generating magnetic fields can be estimated by the modified magnetic
Reynolds number that in dynamo theory is conventionally notated as
$C_\Omega$,
\begin{equation}
    C_\Omega = \frac{\Delta\Omega H^2}{\eta_{_\mathrm{T}}},
    \label{15}
\end{equation}
where $\Delta\Omega$ is the angular velocity variation within the
convection zone, $H$ is the convection zone thickness, and
$\eta_{_\mathrm{T}}$ is the magnetic eddy diffusivity. This number
estimates the ratio of the rate $\Delta\Omega$ of magnetic field
production by differential rotation to the rate
$H^2/\eta_{_\mathrm{T}}$ of diffusive decay of the field.
Fig.~\ref{f9} shows isolines of $C_\Omega$ on the plane of effective
temperature and rotation period, obtained in the same computations
as Fig.~\ref{f8}. $C_\Omega$ increases with rotation rate. This is
mainly due to the rotational quenching of the eddy diffusivity. The
dynamo number (\ref{15}) increases also with decreasing temperature.
This is mainly because convection is slower and $\eta_{_\mathrm{T}}$
is smaller in cooler stars. The increase of $\eta_{_\mathrm{T}}$
with increasing temperature overpowers the increase in differential
rotation, so that the dynamo-number (\ref{15}) decreases.

The large rotational shear of F-stars is less efficient in winding
magnetic fields than the small differential rotation of K-dwarfs.
The widely spread opinion that low mass stars host
$\alpha^2$-dynamos producing non-axisymmetric global fields is not
supported by the estimations of the $C_\Omega$ dynamo-number. The
small but dynamo-efficient differential rotation of these stars
favors axisymmetric global fields. This might be the reason why the
magnetic structure of the M-star observed by \cite[Donati et al.
(2006)]{Dea06} was close to axial symmetry.
 \section{Summary}
We have seen that meridional flow results from (slight) deviations
from the thermal wind balance. The balance, in turn, is maintained
by the flow. Meridional flow attains its largest velocities in the
boundary layers near the top and bottom of the convection zone where
deviations from the thermal wind balance are relatively large. The
thickness of these layers decreases with increasing rotation rate.
The layers in rapidly rotating stars are so thin that an
advection-dominated regime of the dynamo in these stars is not
probable.

Differential rotation is produced mainly by convection and also by
the meridional flow transporting angular momentum. Differential
temperature with poles warmer than the equator is very significant
for differential rotation formation. Differential temperature is
currently understood as a result of rotationally induced anisotropy
of convective heat transport. The anisotropy is the parameter to
which differential rotation models are most sensitive.

Differential rotation models reproduce closely the
helioseismological rotation law and the observed differential
rotation of several individual stars. The models predict the
differential rotation of a star of given mass to vary mildly as the
star ages and its rotation rate decreases. Differential rotation is,
however, predicted to increase strongly with stellar mass. The
significance of rotational shear for dynamos estimated with the
$C_\Omega$ dynamo number (\ref{15}) has the opposite tendency. The
large differential rotation in F-stars is less dynamo-efficient than
the tiny rotation inhomogeneity of M-stars.

\pagebreak[4]

\acknowledgements{}
The author is thankful to the Russian Foundation for Basic Research
(projects 10-02-00148 and 12-02-92691\underline{\ }Ind) and to the
Ministry of Education and Science of the Russian Federation
(contract 16.518.11.7065) for their support.

\end{document}